\documentclass[prl,a4paper,graphicx,preprint,reprint,twocolumn]{revtex4-1}

\usepackage{amsmath,amssymb}
\usepackage{physics}
\usepackage{gensymb}
\usepackage{times}
\usepackage{graphicx,float,wrapfig}
\usepackage{soul}
\usepackage{dsfont}
\pdfoutput=1

\usepackage[normalem]{ulem}

\usepackage{gensymb}

\usepackage{xcolor}
\usepackage[left,switch,mathlines]{lineno}
\modulolinenumbers[2]

\usepackage[colorinlistoftodos]{todonotes}
\usepackage[colorlinks=true, allcolors=blue]{hyperref}

\newcommand{\x}{\mathbf{r}}
\newcommand{\av}[1]{\left\langle #1 \right\rangle}

\renewcommand{\k}{{\bf k}}
\newcommand{\bk}{\mathbf{k}}

\begin{document}

\title{A strong Bell correlation witness between spatially separated pairs of atoms}

\date{\footnotesize \today}

\author{D.~K.~Shin$^{1}$} 
\author{B.~M.~Henson$^{1}$}
\author{S.~S.~Hodgman$^{1}$}
\author{T.~Wasak$^{2}$}
\author{J.~Chwede\'nczuk$^{3}$}
\author{A.~G.~Truscott$^{1}$ }

\email{andrew.truscott@anu.edu.au}

\affiliation{\normalsize{$^{1}$Research School of Physics and Engineering, Australian National University, Canberra 0200, Australia}}
\affiliation{\normalsize{$^{2}$Max Planck Institute for the Physics of Complex Systems, 01187 Dresden, Germany}}
\affiliation{\normalsize{$^{3}$Faculty of Physics,  University of Warsaw,  ul.  Pasteura 5, PL–02–093 Warszawa,  Poland}}

\begin{abstract}
The violation of a Bell inequality is a striking demonstration of how quantum mechanics contradicts local realism \cite{Bell1964}.
Although the original argument was presented with a pair of spin-$\frac{1}{2}$ particles,
so far Bell inequalities have been shown to be violated using entangled pairs of photons \cite{Freedman1972,Aspect1982}, 
with recent measurements closing all possible loopholes in such a scheme \cite{Hensen2015,Giustina2015,Shalm2015}.
Equivalent demonstrations using massive particles have proven to be much more challenging, generally relying on post-selection of data \cite{Lamehi1976,Sakai2006} 
or measuring an entanglement witness that relies on quantum mechanics \cite{Schmied2016}.
Here, we use a collision between two Bose-Einstein condensates to generate the momentum-spin entangled pairs of ultracold helium atoms. We show that a maximally entangled Bell triplet state results and report a direct observation of a strong Bell correlation witness.
Based on the high degree of entanglement and the controllability of ultracold atomic systems, extensions to this scheme would allow a demonstration of nonlocality with massive entangled pairs, 
following Bell's original idea. Other applications include the demonstration of the Einstein-Podolsky-Rosen paradox \cite{EPR1935}, quantum metrology and tests of phenomena from exotic theories sensitive to such systems including gravitational decoherence \cite{Penrose1996} and quantum gravity.
\end{abstract}

\maketitle 

The basic scheme of a Bell test, originally proposed by John Bell \cite{Bell1964}, involves measuring correlated detector events between a pair of spin-$\frac{1}{2}$ particles arriving at spatially separated detectors labelled $A$ and $B$.  Bell envisioned using an entangled singlet state $\ket{\Psi^{-}} = \left(\ket{\uparrow}_A\otimes\ket{\downarrow}_B - \ket{\downarrow}_A\otimes\ket{\uparrow}_B\right)/\sqrt{2}$,
where the states $\ket{\uparrow}$ and $\ket{\downarrow}$ are the spin eigenstates for each particle. 
The particles are subject to independent rotations and then the spins of the pairs are measured.
If the correlation value violates the so-called Bell inequality, then any description of the system will be incompatible with local realism, which requires the particles to be in a defined state at all times and not to communicate faster than the speed of light.
In quantum mechanics, the particles form an entangled state and only occupy a definite state once the measurement occurs.  
Entanglement forms the basis for a range of emerging and future quantum technologies, such as quantum computing or metrology \cite{Horodecki2009}.

In Bell tests with photons, the non-classical state comes from parametric down-conversion.  The photons are entangled in polarisation (analogous to spin), which is rotated using waveplates and then measured.  While the earliest experiments with photons \cite{Freedman1972,Aspect1982} contained several loopholes that still allowed for interpretations consistent with local realism, more recent experiments have succeeded in closing all loopholes \cite{Hensen2015,Giustina2015,Shalm2015}.  The first experiments using massive particles were performed by measuring spin correlations between high energy particles \cite{Lamehi1976,Sakai2006}, while a more recent loophole-free experiment looked at entanglement between the spins in solid state systems \cite{Hensen2015}.  However, such schemes rely on entanglement between internal degrees of freedom, and are thus unable to be extended to entanglement between motional degrees of freedom. This is a major motivation for looking at entanglement between atoms, as spatially separated, motional entanglement could provide insights into proposed theories of quantum gravity, by measuring their gravitational decoherence \cite{Penrose1996}.

One promising experimental system for demonstrating such entanglement is ultracold atoms \cite{Lewis-Swan2015,Wasak2018}.
Indeed, many characteristic quantum effects have been realised recently in ultracold atomic systems ranging from the generation of non-classical atomic pairs \cite{Kheruntsyan2012}, Hong-Ou-Mandel interference \cite{Lopes2015}, and the observation of spatially separated entanglement \cite{Lange2018} and Einstein-Podolsky-Rosen (EPR) steering \cite{Fadel2018,Kunkel2018} in an ensemble. 
A quantum mechanical (QM) witness of many-body Bell correlations has also been observed in the collective spin of a Bose-Einstein condensate (BEC) \cite{Schmied2016}. 
However, the spins were not spatially separated, which makes it challenging to extend such a scheme to demonstrate nonlocality.
Promising progress on a test for momentum entanglement in an atomic pair has also been reported \cite{Dussarrat2017}.

\begin{figure}[ht]
\centering
\includegraphics[width=8cm]{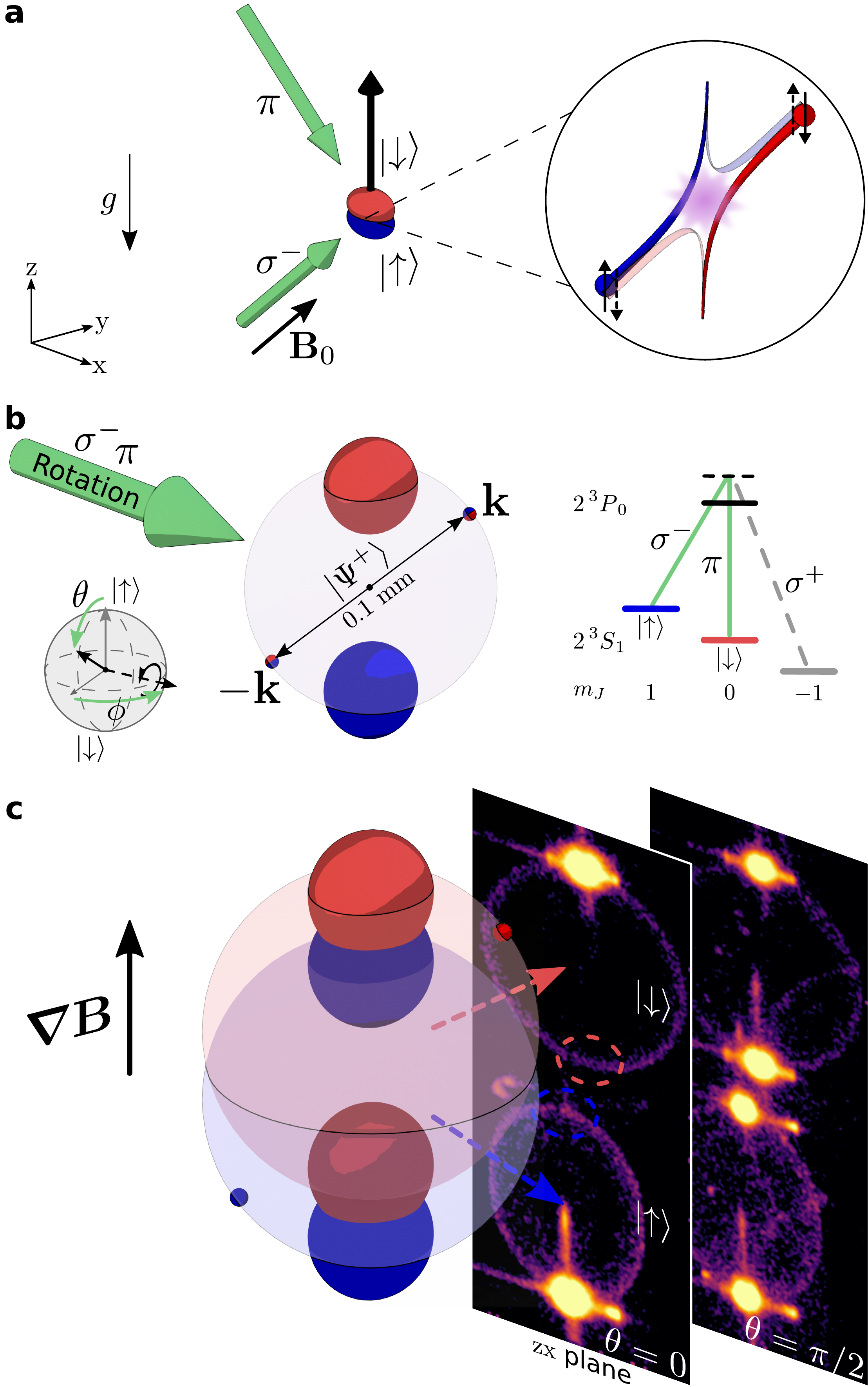} 
\caption{\label{fig:exp_schem}
Experimental schematic for the generation and detection of entangled pairs of atoms.
(a) Binary collisions entangle pairs of particles in quantum mechanics due to the symmetry in certain wavefunctions (inset).
In this work, such entanglement is realised in pairs of atoms from the collision products of oppositely spin-polarised BECs.
(b) The scattered pairs form a spherical shell as antipodal points in momentum, where the BECs lie on the two poles along the collision axis.
Pairs spatially separate in time at which point individual atom's spin is rotated by an angle $\theta$.
(c) An applied magnetic field gradient spatially separates the atoms by spin, before individual atoms are detected with full 3D momentum and spin resolution. The images on the right show the atom count density in the $zx$-plane for two values of $\theta$.
Spin correlations between the scattered pairs (diametrically opposite regions on the purple rings) will exhibit quantum nonlocality.
The bright yellow ellipses correspond to the BECs, which saturate the detector.
}
\end{figure}

In this paper, we report on the generation and detection of a strong Bell correlation witness between the atomic spins across a spatially separated pair of metastable helium (He*) atoms.
The experiment consists of the three essential components necessary to realise a Bell test: a correlated atomic pair source, a rotation of the spins of both atoms corresponding to an independently configurable measurement basis and the momentum and spin resolved single-particle detection necessary for evaluating pair correlations.  Each stage is described in detail below and shown schematically in Figure~\ref{fig:exp_schem}.
Briefly, the pair source is the binary scattering product from a collision between two oppositely spin-polarised BECs (Fig.~\ref{fig:exp_schem}a) which naturally separate in time. 
The spins of both atoms in each pair are then rotated by the same angle (Fig.~\ref{fig:exp_schem}b) followed by a direct measurement of their momentum and spin (Fig.~\ref{fig:exp_schem}c). 
Over many experimental runs we extract the correlations between the spins, which show excellent agreement with the predictions of quantum mechanics, and witness the Bell correlations in our system. We experimentally demonstrate this Bell correlation witness with a significant spatial separation of $\approx0.1~\textrm{mm}$ across the entangled pairs of atoms, and determine the form of quantum entanglement using a quantum state tomography technique operationally akin to the Bell test.
By experimentally observing a strong Bell correlation witness for a range of rotation angles, the quantum state of the pairs is demonstrated to be a Bell triplet state, suitable for showing quantum nonlocality in a more general Bell test. 

Our experiment starts with a magnetically trapped BEC of helium-4 atoms in the long-lived metastable state $2\,{}^3S_1$ (see Methods for details).  The atomic sublevels $\ket{J=1,m_J=1}=\ket{\uparrow}$ and $\ket{J=1,m_J=0}=\ket{\downarrow}$ form the qubit subspace (see level diagram in Fig.~\ref{fig:exp_schem}(b)), with the atoms initially fully spin-polarised in the $\ket{\uparrow}$ state.
Following trap switch-off, a $\pi/2$-pulse from a two-photon stimulated Raman process via the $\lambda=1083~\mathrm{nm}$ $2\,{}^{3}S_{1} - 2\,{}^{3}P_{0}$ transition (see Fig.~\ref{fig:exp_schem}(b)) then simultaneously flips half of the atoms' spin to $\ket{\downarrow}$ and imparts a velocity of $\sim 120~\mathrm{mm/s}$ along the $z$-axis, opposite to gravity (see Fig.~\ref{fig:exp_schem}(a)).
In the centre of mass frame, the two condensates split apart at $v_{\mathrm{r}} \approx \pm60~\mathrm{mm/s}$ and spontaneously scatter atoms into correlated pairs of opposite momenta and spin via binary elastic $s$-wave scattering, forming a uniformly distributed spherical halo in momentum with radius $k_r=2\pi/\sqrt{2}\lambda$ \cite{Perrin2007,Khakimov2016}.
The opposite spin-states of the colliding BECs entangle the pairs in spin as well as momentum (see inset of Fig.~\ref{fig:exp_schem}(a)).
With the momenta of each pair given by $(\mathbf{k},-\mathbf{k} = A,B)$, the state is symmetric under exchange of labelling by momentum and exhibits complete anti-correlation in spin. 
Bogoliubov scattering theory predicts that the state of the pair is the archetypal Bell triplet $\ket{\Psi^+} = \left(\ket{\uparrow}_{A}\otimes\ket{\downarrow}_{B} + \ket{\downarrow}_{A}\otimes\ket{\uparrow}_{B}\right)/\sqrt{2}$ (see Methods for details).
Such a state is maximally entangled, useful in various quantum information tasks \cite{Horodecki2009}, and, more importantly to this work, a viable candidate for demonstrating nonlocality \cite{Wiseman2007}.

Following the collision pulse, the scattering halo evolves freely in a uniform magnetic field of $\sim0.5~\mathrm{G}$ for $t_{\mathrm{sep}}=0.8~\textrm{ms}$ (see Fig.~\ref{fig:exp_schem}(b)).
The halo expands spherically such that each entangled pair, located at diametrically opposite regions of the halo, is spatially separated by $d_{\mathrm{sep}} \approx 0.1~\mathrm{mm}$.
A pair of co-propagating Raman beams (see Fig.~\ref{fig:exp_schem}(b)) that are wider than the size of the halo by over an order of magnitude provide a uniform rotation corresponding to $\hat{R}_{y}(\theta)=\exp\left(-i\theta\hat\sigma^{(A)}_y\right)\otimes\exp\left(-i\theta\hat\sigma^{(B)}_y\right)$, where $\hat\sigma^{(A)}_y$ and $\hat\sigma^{(B)}_y$ 
represent the $y$-component of Pauli matrices for
spins at $A$ and $B$, while imparting no net momentum change to the atoms. The rotation is independent of the atom's momentum and position, is applied to the whole atomic ensemble, with the rotation angle $\theta$ controlled by the optical pulse duration.  A key feature of the $\ket{\Psi^+}$ state is that it is not rotationally invariant under a uniform rotation of both atoms in the pair by a single angle $\theta$, which enables us to measure the entanglement of the state.  

Immediately after the rotation pulse, a magnetic field gradient is applied in the $z$-direction (see Fig.~\ref{fig:exp_schem}(c)).
This projects the atoms into the $\hat{S}_z$ eigenstates $\left\lbrace\ket{\uparrow},\ket{\downarrow}\right\rbrace$ via the Stern-Gerlach effect, separating the two spin states in the vertical direction.
Since only the $m_J=1$ state has a non-zero magnetic moment, only $\ket{\uparrow}$ feels a magnetic force, causing the state to spatially separate from $\ket{\downarrow}$ atoms at the detector and allowing state-resolved detection.  

The atoms then freely fall under gravity onto the detector located $0.848~\textrm{m}$ below, from which we obtain the momentum and spin information for individual He* atoms.
Figure~\ref{fig:exp_schem}(c) shows a typical image from an average of many experimental shots, displaying two completely separated halos for each of the experimental configurations in which there was no rotation pulse and a $\pi/2$-rotation that evenly mixes spins.
Atoms in the $\ket{\uparrow}$ state form the lower halo, which is slightly non-spherical due to inhomogeneity in the magnetic field gradient causing a spatially dependent force around the halo (see Fig.~\ref{fig:exp_schem}(c)).
Such distortion corresponds to a misalignment of the ideal back-to-back pairing in momentum and is removed in the data analysis (see Methods for details).
Since the $m_J=0$ states are unaffected by magnetic fields, the $\ket{\downarrow}$-halo maintains the $s$-wave spherical shell shape at the detector (see the upper halo in Fig.~\ref{fig:exp_schem}(c)).

\begin{figure}[ht]
\centering
\includegraphics[width=8.5cm]{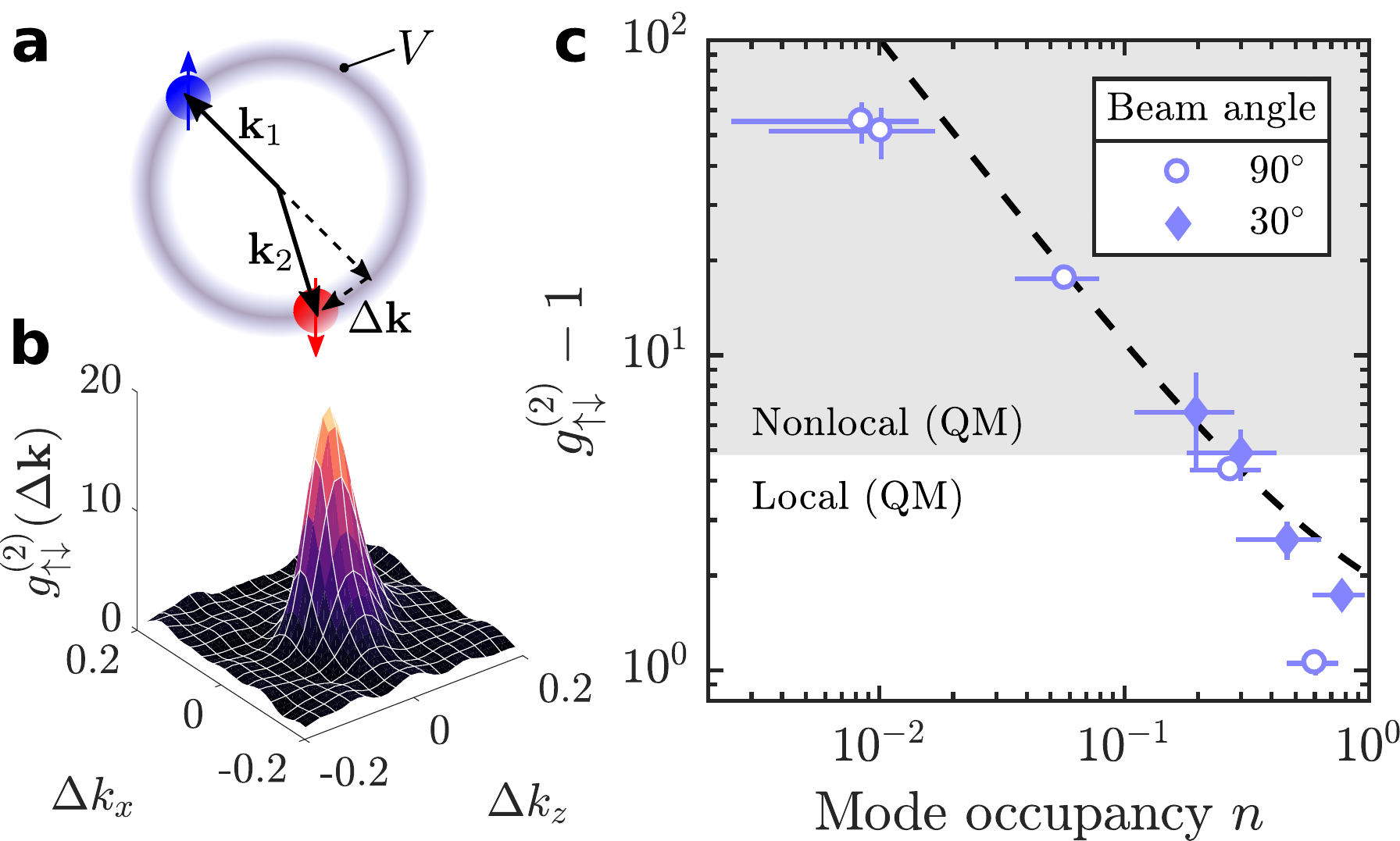}
\caption{\label{fig:fig_src-g2}
Tunable momentum-spin anti-correlated atomic pair source.
(a) Schematic of atoms with momentum and spin degrees of freedom in the scattering sphere.
A 2D planar slice in momentum space is taken for simplicity.
(b) Two-body cross-correlation function in momentum-spin $g^{(2)}_{\uparrow\downarrow}$ from the un-rotated pair source with an average mode occupation of $n=0.058(2)$, averaged over 2,100 experimental runs.
(c)The dependence of the degree of anti-correlation on the average halo mode occupation.  Two different angles between the Raman beams ($30{\degree}$ and $90\degree$) were used to produce the experimental data.
The dashed line depicts the theoretical prediction.
Demonstration of nonlocality strictly requires a minimum correlation strength in the pair source of $g^{(2)}_{\uparrow\downarrow}> 3+2 \sqrt{2}$.
All error bars indicate the standard deviation in the mean (see Methods for details).
}
\end{figure}

To characterise the two-state scattering halo we look at two-particle correlations between atoms on opposite sides of the halo with either parallel or anti-parallel spin-pairing, given by
\begin{equation}
g^{(2)}_{
ij} (\Delta \mathbf{k}) = \frac{\sum\nolimits_{\mathbf{k} \in V} \left\langle \hat{n}_{\mathbf{k},i} \hat{n}_{-\mathbf{k} + \Delta\mathbf{k},j} \right\rangle}{\sum\nolimits_{\mathbf{k} \in V}  \left\langle \hat{n}_{\mathbf{k},i} \right\rangle\left\langle \hat{n}_{-\mathbf{k} + \Delta\mathbf{k},j} \right\rangle},
\label{eqn:g2}
\end{equation}
where $i, j\in \{\uparrow,\downarrow\}$ denote spin states, $\hat{n}_{\mathbf{q}, m}$ the number of atoms with momentum $\mathbf{q}$ and spin $m$, and $V$ the volume in momentum space occupied by the $s$-wave scattering halo \cite{Hodgman2017} (shown schematically in Fig.~\ref{fig:fig_src-g2}(a)).
Figure~\ref{fig:fig_src-g2}(b) shows an experimentally measured correlation function for scattering halo with no rotation, which displays a large peak indicative of strong correlations between atoms in different spin states on opposite sides of the halo. 
The pair correlation amplitude is inversely proportional to the mode occupancy $n$ for a spontaneous pair source \cite{Hodgman2017}, which we tune by varying the number of atoms in the colliding BECs. 
This relationship is verified in Fig.~\ref{fig:fig_src-g2}(c), confirming that our pair source behaves as expected. Furthermore, the inverse proportionality is consistent with the predictions of Bogoliubov theory, which describes the pair-scattering process in the low-gain regime. Importantly, we are able to reach correlation amplitudes of $\sim60$, although due to signal to noise considerations we actually operate in a regime of $g^{(2)}_{\uparrow \downarrow}(0) \approx 30$, an amplitude sufficient to demonstrate a violation of a Bell inequality~\cite{Wasak2018}. The corresponding average mode occupation in the scattering halo of $\sim0.03$ means we are operating in the low-gain regime, where the dominant contribution to the halo comes from scattering of single pairs.

\begin{figure}[htb]
\centering
\includegraphics[width=8cm]{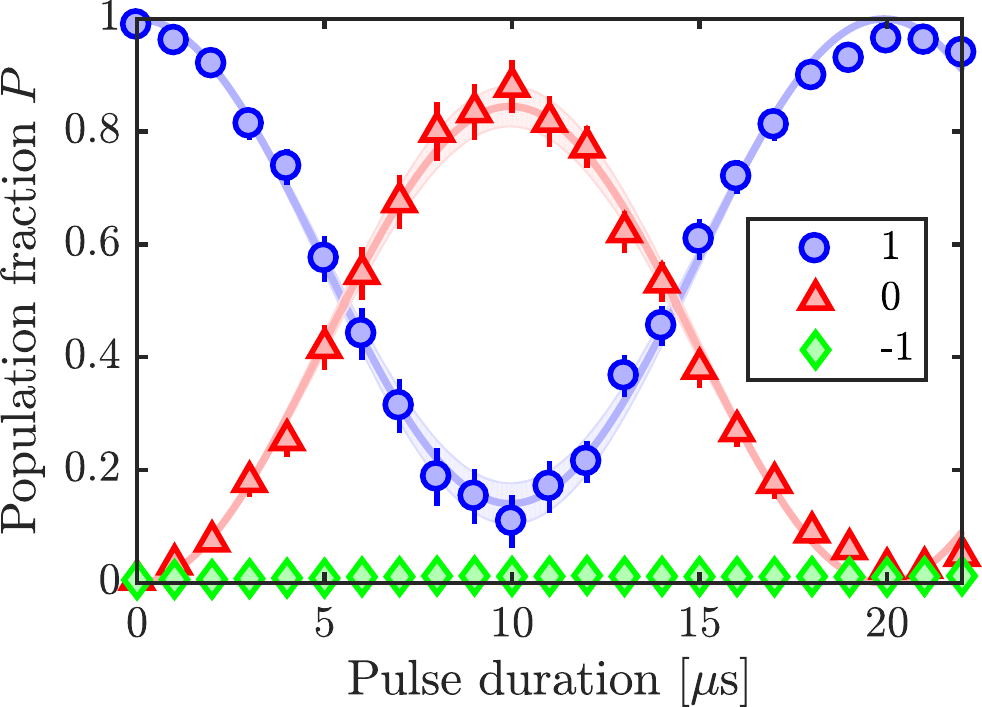}
\caption{\label{fig:fig_gate-char}
Coherent control of atomic spin in a metastable helium scattering halo.
Rabi oscillation in the population fraction from a single rotation pulse with an amplitude of $0.85(4)$ and effective Rabi frequency $\Omega' = 2\pi\cdot50.3(3)~\mathrm{kHz}$.
The Larmor precession frequency of the atomic spins is $\Omega_L \approx 2\pi\cdot1.4~\mathrm{MHz}$ for a field of $\sim0.5~\mathrm{G}$.
The shaded region around the fitted model (solid line) indicates variations in the Rabi oscillation characteristics at different momentum zones of the scattering sphere.
}
\end{figure}

To characterise the spin rotation a $\ket{\uparrow}$-polarised scattering halo was initially prepared from a Raman sequence 
similar to our previous work to generate $m_J=0$ halos (see \cite{Khakimov2016,Hodgman2017} and Methods).  The atoms are then rotated by $\hat{R}_y(\theta)$ and the Rabi oscillations are observed (see Fig.~\ref{fig:fig_gate-char}) with negligible coupling to the $m_J = -1$ state.

\begin{figure*}[tp]
\centering
\includegraphics[width=15cm]{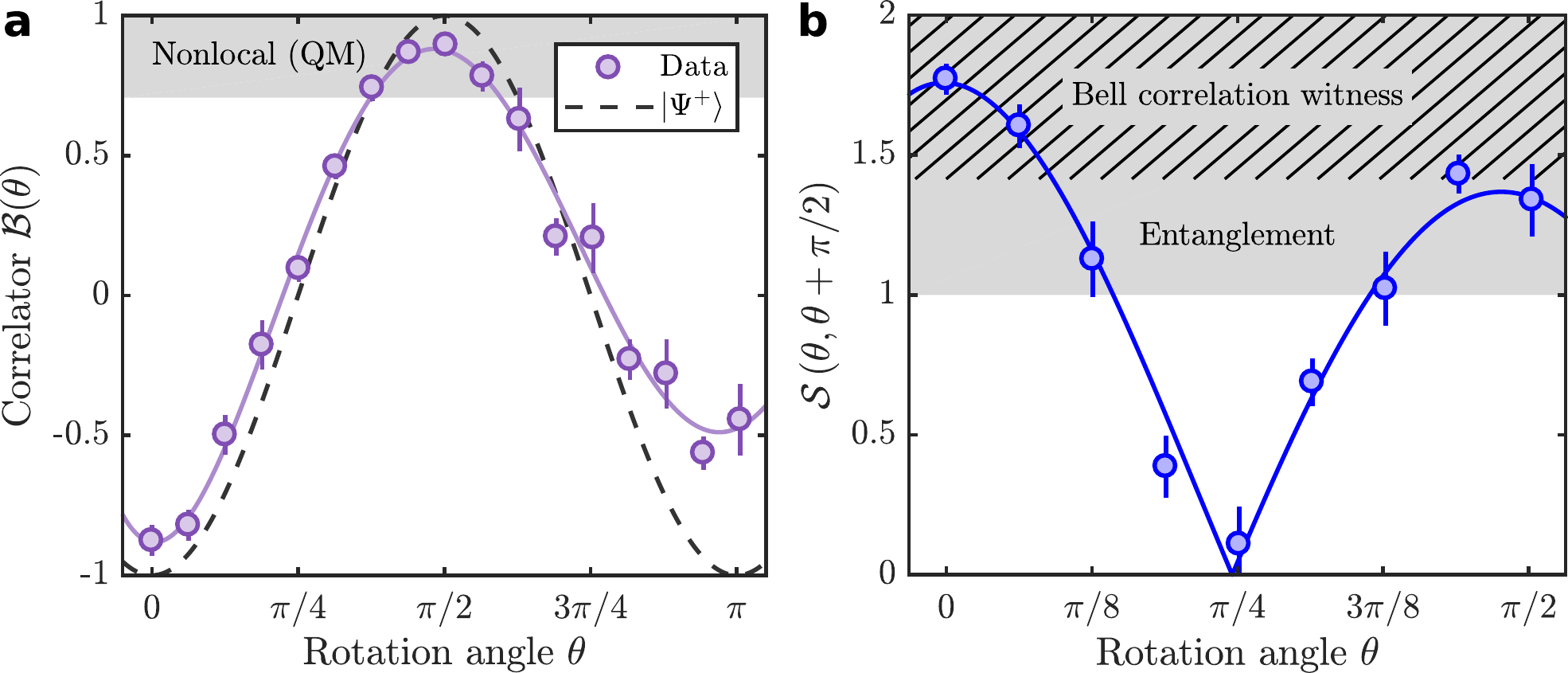}
\caption{\label{fig:fig_bell} 
Non-classical correlations in scattering halos.
(a) A strong correlation $\mathcal{B}$, exceeding $1/\sqrt{2}$ (shaded region), signals the potential to observe the violation of a Bell inequality based on a quantum mechanical model. 
The QM prediction of the violation was observed by $\mathcal{B}(\pi/2) = 0.90(1)$ with a significance of $18$ standard deviations from 18,000 experimental runs when the spins were rotated by $\pi/2$.
The dashed line is the theoretical prediction for the Bell triplet $\ket{\Psi^+}$ and the solid line is a guide to the eye.
(b) Direct observation of non-classicality in the pair correlations based on the QM witness of Bell correlations $\mathcal{S}$.
Correlations lying in the shaded and hatched regions indicate the presence of quantum entanglement and Bell nonlocality, respectively.
6-sigma violation of the Bell witness inequality \eqref{eqn:Bell_nonlocal} is observed by $\mathcal{S}(0,\pi/2) =1.77(6)$
All error bars correspond to the standard error evaluated from bootstrapping (see Methods for details).
}
\end{figure*}

By rotating the dual spin halo and measuring the resulting correlations between atoms in each state (see Methods for details), a correlator
\begin{equation}\label{eq.corr.1}
	\mathcal B(\theta)=\av{\hat\sigma_z^{(A)}\hat\sigma_z^{(B)}}_\theta=\frac{g^{(2)}_{\uparrow\uparrow}+g^{(2)}_{\downarrow\downarrow}-g^{(2)}_{\uparrow\downarrow}-g^{(2)}_{\downarrow\uparrow}}
{g^{(2)}_{\uparrow\uparrow}+g^{(2)}_{\downarrow\downarrow}+g^{(2)}_{\uparrow\downarrow}+^{(2)}_{\downarrow\uparrow}}
\end{equation}
is obtained (the $\theta$ subscript denotes the average in the rotated state). This correlator is displayed in Fig.~\ref{fig:fig_bell}(a), with the experimental results showing excellent agreement with the theoretical prediction
$\mathcal B(\theta)=-\cos2\theta$ for the Bell triplet state 
$\ket{\Psi^+}$. This is the first strong indication that the two atoms are strongly entangled. 

To prove the non-classical properties of our two-atom system, we first show that the pairs are entangled. Note that for all non-entangled states, the maximum range of the correlator (\ref{eq.corr.1}) is bounded by unity
\begin{equation}
	\mathcal{S}(\theta,\theta') = |\mathcal B(\theta)-\mathcal B(\theta')|\leqslant1
	\label{eqn:general_Sparam}
\end{equation}
(see Methods for details).
We detect a clear violation of this bound in Fig.~\ref{fig:fig_bell}(a), 
which proves the system is entangled --- a necessary ingredient for the violation of a Bell inequality for any quantum system.

Since we have shown that the atomic spins are vector quantities under rotation (see Figure~\ref{fig:fig_gate-char} and Methods), we can now exclude a wide class of local hidden variable (LHV) theories.  Violation of the inequality
\begin{equation}
	\mathcal S\left(\theta,\theta + \frac{\pi}{2}\right)=\left|\mathcal B(\theta)-\mathcal B\left(\theta+\frac\pi2\right)\right|\leqslant\sqrt2,
	\label{eqn:Bell_nonlocal}
\end{equation}
from two complementary measurements certifies the exclusion of situations in which one subsystem gives binary outcomes, whereas the second consists of a vector quantity (see Methods for details).
This is violated in our system, as shown in Fig.~\ref{fig:fig_bell}(b), since for $\theta=0$ we observe $\mathcal{S}(0,\pi/2) =1.77(6)$. 

Finally, we compare our results to the predictions of Bogoliubov theory applied to the scattering process (see Methods for details), which has been tested for a wide range of pair production processes, ranging from the parametric down conversion of photons to the emission of entangled atoms from colliding BECs (as in our case). 
In particular, this theory predicts that $\mathcal B(\theta)=-\cos2\theta$. Furthermore the Bogoliubov model provides a necessary condition for the future observation of the violation of the Clauser, Horne, Shimony and Holt (CHSH) inequality in our system \cite{clauser1969proposed}, namely $|\mathcal B(\theta)|>\frac1{\sqrt2}$ (see Methods for details). These are both confirmed by the experimental data in Fig.~\ref{fig:fig_bell}(a).

In conclusion, we have demonstrated the creation and coherent control of maximally entangled pairs of He* atoms, obtained from an $s$-wave collision halo generated from BECs in two different spin states. 
We have characterised the correlations between the pairs and have shown that they are sufficiently strong to exhibit Bell correlations and to demonstrate the nonlocality in this particular Bell test.
This was demonstrated by rotating the spin of each atom by a variable angle using a Raman transition.
The spin and momentum resolved single-atom detection is implemented by applying a magnetic field gradient during time-of-flight.
This yields a strength of the correlations exceeding the bound needed to violate the Bell inequality by 18$\sigma$.
Extensions to this scheme will allow the demonstration of EPR steering, quantum metrology, and coherent control of the He* qutrit in $2{}^3S_1$, as well as potential applications for the entangled pairs in quantum technologies including atom interferometry and quantum information. 
In addition, by implementing independent rotations at different parts of the halo, a Bell test almost identical to the original proposal \cite{Bell1964} and its modification ~\cite{clauser1969proposed} is possible.
On a more fundamental level, this demonstration of a strong degree of entanglement between massive particles separated over a macroscopic distance opens the door to testing exotic theories of gravitational decoherence \cite{Penrose1996}.

\begin{acknowledgments}
 The authors would like to thank Alain Aspect, Michael Barson, Danny Cocks, Matteo Fadel, Karen Kheruntsyan, Wolfgang Mittig, Margaret Reid, Jacob Ross and Tilman Zibold for insightful discussions. 
 This work was supported through Australian Research Council (ARC) Discovery Project grants DP120101390, DP140101763 and DP160102337. 
DKS is supported by an Australian Government Research Training Program Scholarship.
 SSH is supported by ARC Discovery Early Career Researcher Award DE150100315.  
 JC is supported by Project no. 2017/25/Z/ST2/03039, funded by the National Science Centre, Poland, under the QuantERA programme.
\end{acknowledgments}

\bibliographystyle{unsrt}

\section*{Methods}

\subsubsection*{Experimental apparatus and procedure}
The He$^*$ BEC is initially prepared in the $m_J=1$ state in a bi-planar quadrupole Ioffe configuration magnetic trap as explained in our previous papers~\cite{Dall2007,Khakimov2016}, with harmonic frequencies of $(\omega_x, \omega_y, \omega_z)/2\pi\approx(15,~25,~25)$~Hz.
The magnetic trap is switched-off abruptly, here denoted as time $t=0$, from which it takes $\sim2~\textrm{ms}$ for the magnetic field to stabilise to a uniform field of $\mathbf{B}_0 \approx0.5\left(\mathbf{e}_x + \mathbf{e}_z\right)/\sqrt{2}~\textrm{G}$, which splits the degeneracy in spin by $f_{\updownarrow}=g\mu_{0}B\approx1.4~\mathrm{MHz}$ and is maintained throughout until the Stern-Gerlach sequence.

At $t=3~\textrm{ms}$ the $\pi/2$ collision Raman pulse for creating $\ket{\Psi^+}$-pairs, lasting $\sim10~\mu\textrm{s}$, is applied from two $90^{\circ}$-crossed laser beams $L_1$/$L_2$, propagating along the $(\mathbf{e}_x\pm\mathbf{e}_z)/\sqrt{2}$ directions, and $\sigma^-$/$\pi$- polarised with respect to the quantisation axis defined by $\mathbf{B}_0$, respectively.  
Each beam's optical frequencies were far-detuned from the $2\,{}^{3}S_{1} - 2\,{}^{3}P_{0}$ transition by $\Delta\approx 3~\mathrm{GHz}$ such that $\Delta/\Gamma \approx 2000 \gg 1$, making the spontaneous single photon absorption rate negligible.
A single photon recoil is $\hbar k_0=2\pi\hbar/\lambda$, where $\lambda=1083.20~\mathrm{nm}$.

The atoms evolve freely in the stabilised magnetic field $\mathbf{B}_0$ for $0.8~\textrm{ms}$, at which point the scattering halo, uniformly expanding at a rate $\dot{d}_{\mathrm{sep}}\approx 120~\textrm{mm/s}$ (given by the recoil momenta from the two photons absorbed by the He* atoms), reaching a diameter of $d_{\mathrm{sep}}\approx96~\mathrm{\mu m}$.
The spin rotation pulse is then applied at $t=3.8~\textrm{ms}$, from a second stimulated Raman transition coupled to the same transition and detuned as above, using a single beam $L_3$ that propagates along the $x$-axis.
An RF-pulsed acousto-optic modulator produces a two-tones optical pulse in $L_3$, with the two frequencies fulfilling the co-propagating resonance condition for the two photon Raman process.
Furthermore, the beam was elliptically polarised with propagation along $x$-axis such that $\sigma^+$ polarisation was extinguished along $\mathbf{B}_0$, which would otherwise couple the qubit subspace to the $m_J=-1$ substate.
The beam waist at the trap was $\sigma_3\approx1.1~\textrm{mm}$, an order of magnitude larger than spatial extent of the atomic ensemble at the time of rotation sequence, which provided a uniform rotation operation for all scattered atoms in the halo.

After the rotation pulse, the Stern-Gerlach sequence is implemented by pulsing current through a large coil concentric to the $z$-axis, that selectively pushes the $m_J=+1$ atoms along the $-z$ direction.
Subsequently, the atoms are detected by an $80~\textrm{mm}$ diameter microchannel plate and delay line detector located $848~\textrm{mm}$ below the trap.  The free-fall duration gives the time at detection $t\approx416~\textrm{ms}$, while the detector has a spatio-temporal resolution of approximately $120~{\mu \mathrm{m}} \times 120~{\mu \mathrm{m}} \times 3~{\mu \mathrm{m}}$ \cite{Henson2018} and a quantum efficiency of $\sim10\%$.

\begin{figure}[htb]
\centering
\includegraphics[width=8cm]{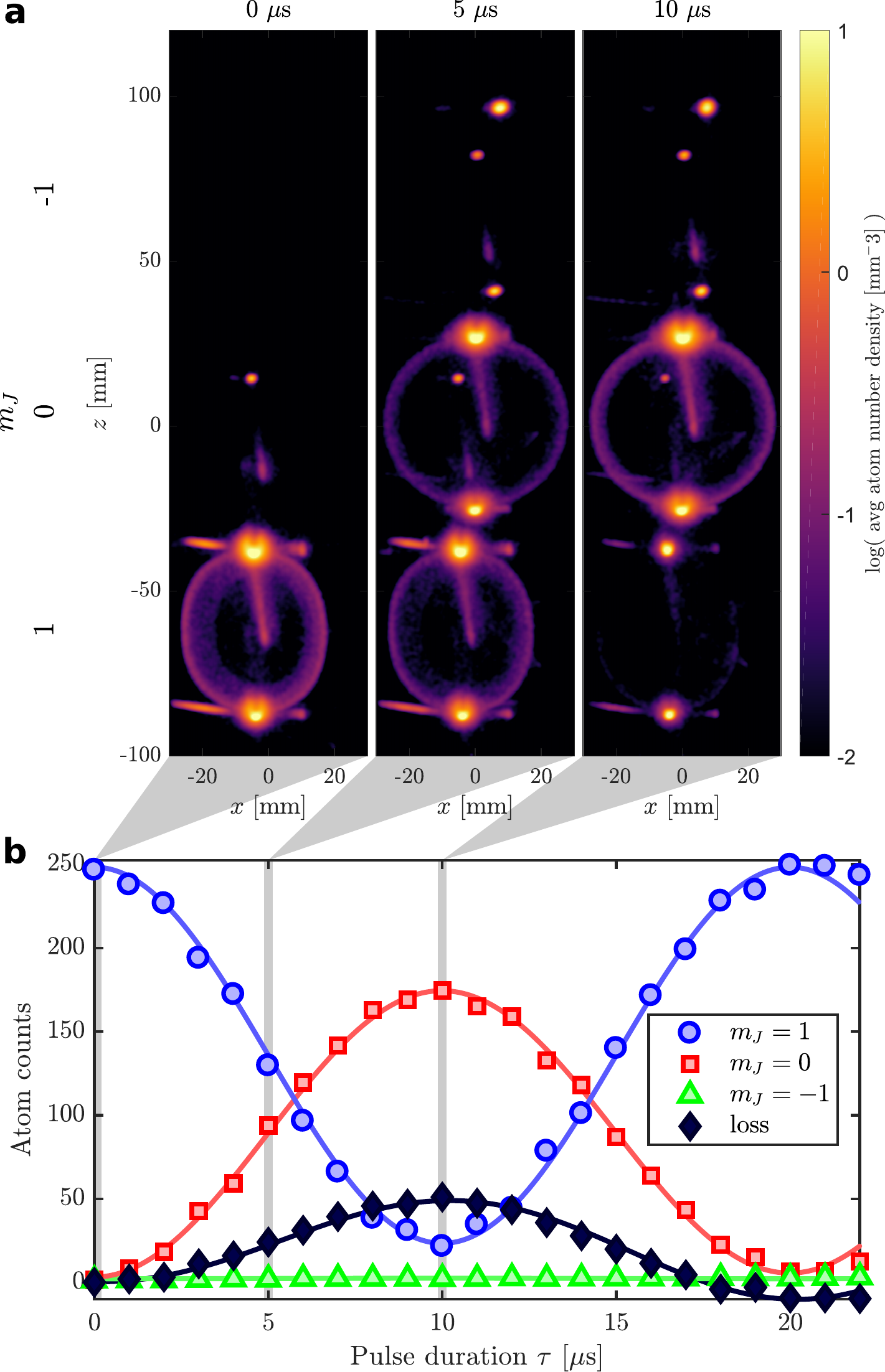}
\caption{\label{fig:sfig_num_mix}
Characterisation of the rotation pulse.
(a) Atom count density at the detector in the $zx$-plane, integrated over $-12~\textrm{mm} < y < 15~\textrm{mm}$, for different spin rotation pulses applied to $\ket{\uparrow}$-polarised scattering halo.
(b) Atom counts in each scattering halo.
Significant loss in total number is observed (black diamond), which is most likely due to Penning ionisation based on the strong correlation with population in $m_J=0$.
For longer pulse durations the atom number in the scattering volume increases due to spontaneous absorption of the Raman beams by the BECs.
Solid lines are sine curve fits to data.
}
\end{figure}

\begin{figure}[htb]
\centering
\includegraphics[width=8cm]{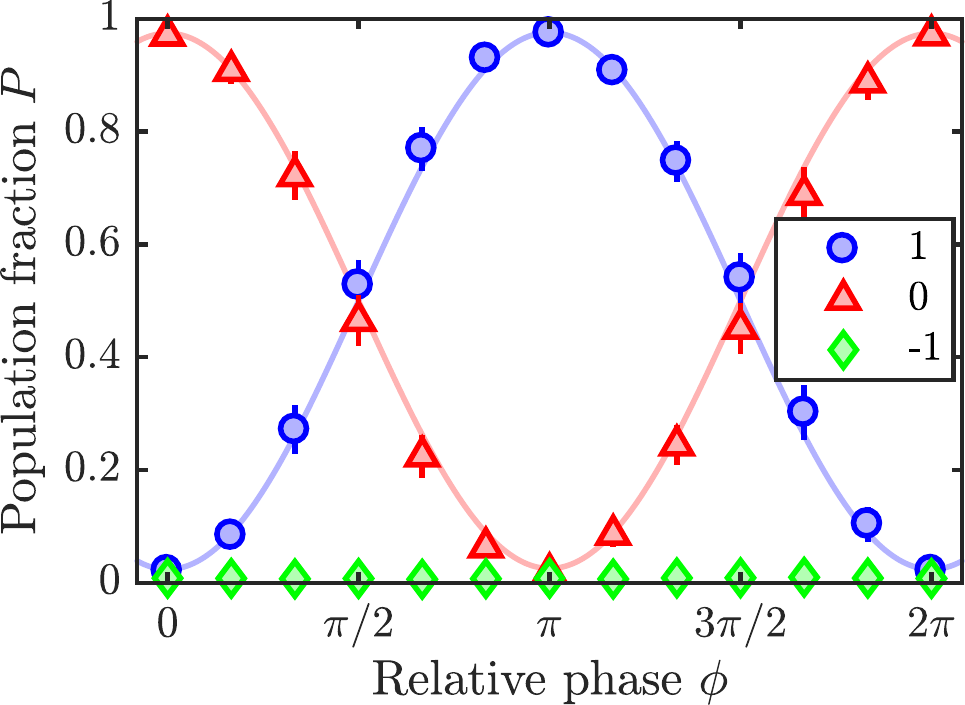}
\caption{\label{fig:sfig_ramsey}
Ramsey-type fringe from two separated $\pi/2$-pulses with a phase delay in the second rotation pulse.
The large fringe visibility of $0.95(2)$ demonstrates the ability to independently configure the axis of rotation ($\mathbf{n}$) around the $xy$-plane of the Bloch sphere.
}
\end{figure}

\subsubsection*{Characterisation of spin rotation}

This section provides the experimental details used to obtain the Rabi oscillation in Fig.~\ref{fig:fig_gate-char}, from which we concluded that the experimental operation in Fig.~\ref{fig:exp_schem}(b) corresponds to a coherent rotation of the qubit.
The characterisation follows a sequence almost identical to the main experiment (Fig.~\ref{fig:exp_schem}), except that only $m_J = +1$ states are prepared initially (see circuit diagram in Fig.~\ref{fig:fig_gate-char}(a)).
Here the collision pulse is applied at $t\approx0$ so that a Bragg transition (no change in internal state) can be driven with the same beam geometry used to produce the entangled pairs (see Fig.~\ref{fig:exp_schem}(a)), producing a $\ket{\uparrow}$-polarised scattering halo (see left Fig.~\ref{fig:sfig_num_mix}(a)).
At the time and location of the Bragg collision pulse, the magnetic field points along the $x$-axis similar to \cite{Khakimov2016}, after which it is stabilised to $\mathbf{B}_0$ for the rotation pulse as described in the previous section.
The rotation pulse is applied at $t=3.8~\textrm{ms}$ so that the halo is $d_{\mathrm{sep}}\approx0.46~\textrm{mm}$ in diameter, approximately five times larger than the $\ket{\Psi^+}$-halo at the point of rotation, but still significantly smaller than the beam waist.

The resulting atom number density at the detector is shown in Fig.~\ref{fig:sfig_num_mix}(a) for various durations of rotation pulse $\tau$.
The total number of atoms detected in the truncated scattering volume (details are given in the next section) $N'_{\alpha}$ is shown in Fig.~\ref{fig:sfig_num_mix}(b) for various rotation pulse durations.
By defining the population fraction of the internal state $\alpha=\pm1,0$, by $P(\alpha) = N'_{\alpha}/N'$, where $N' = \sum_{\alpha} N'_{\alpha}$ is the sum over the triplet, we obtain the Rabi oscillation shown in Fig.~\ref{fig:fig_gate-char}(a).
Although negligible transfer to $m_J = -1$ was achieved, we noticed a loss in the total number of atoms detected in the scattering volume, $N'_{\textrm{loss}}(\tau) = N'(0) - N'(\tau)$ (see Fig.~\ref{fig:sfig_num_mix}(b)).
Two key features evident in the behaviour of the total atom number during the Raman transition are the loss strongly correlated with the presence of $m_J=0$ states and the steady, but weak increase of the total population, which we explain below.

The significant decrease in the total atom number, observed to be up to $25\%$ and correlated with the $N'_0$ population (see Fig.~\ref{fig:sfig_num_mix}(b)), is due to Penning ionisation between pairs of He* atoms \cite{Vassen2012},
which is enhanced by 4 orders of magnitude for spin unpolarised pairs, compared to pairs of 
$m_J=+1$ atoms.

In our experiment, Penning ionisation would most strongly affect the entangled pairs at the earliest time following their production, since the local density of atoms is the highest during the spatial overlap of the pairs with the BECs, allowing for more frequent Penning ionising collisions.
After the BEC and entangled pair wavefunctions have spatially separated, there should be a negligible fraction of atoms lost, since the $\ket{\Psi^+}$ pair source was prepared with very low numbers - an order of in magnitude lower than the scattering halo used to characterise the Raman pulse in Fig.~\ref{fig:sfig_num_mix}.
In the presence of BECs in both spin-states, each atom from the $\ket{\Psi^+}$ pair is almost equally likely to be lost by Penning ionisation.
The loss of an atom from any pair results in the detection of a single-hit event,
i.e. a single-hit event will be detected in $A$ with $\uparrow$ or $\downarrow$, with no correlated hit in $B$.  Such events are thus naturally treated by the correlation functions $g^{(2)}_{ij}$ as an uncorrelated background event (see Eq.~(\ref{eqn:g2})).
The presence of Penning ionisation will therefore strictly can only reduce the observed correlator to the asymptotic uncorrelated state $\mathcal{B}=0$, such that in the extreme case where half of every pair is lost, $g^{(2)}_{ij}=1$ for all back-to-back correlations will be observed.

The increase in total number ($N'_{\textrm{loss}}<0$) for longer pulse durations ($\tau > 17~\mu\textrm{s}$) is due to the constant spontaneous absorption of the Raman beams and subsequent decay by the BECs ($2\,{}^{3}S_{1} \rightarrow 2\,{}^{3}P_J \rightarrow 2\,{}^{3}S_{1}$ where each process is accompanied with a single photon recoil),
into a scattering volume in momentum space intersecting the $s$-wave scattering halo.
The entangled atoms of interest originally occupying the halos are indeed equally subject to such spontaneous processes, and contribute to an additional loss term, negligible in comparison to that of BECs due to the relative atom numbers involved. 
In our experiment, a large detuning of the Raman beams from resonance minimised the rate of spontaneous absorption, as demonstrated by the nearly absent spontaneous effects even from the BECs.

\subsubsection*{Coherent Rotation}
As shown in Fig.~\ref{fig:fig_gate-char}, a Rabi oscillation through $2\pi$-rotation can be induced between the $\ket{\uparrow}$ and $\ket{\downarrow}$ states, with negligible coupling to the $m_J = -1$ state.
To ensure that the Raman pulse is producing the desired rotation of the atomic spin on the Bloch sphere, we implement a Ramsey-type interferometry where a secondary $\pi/2$-pulse with a phase delay $\phi$ follows the first $\pi/2$-pulse $\hat{R}_{y}(\pi/2)$.
Following an identical initialisation scheme to the Rabi oscillation experiment, the fixed delay between the two $\pi/2$-pulses was set at $T \approx 10 \cdot T_\mathrm{L}$, and the phase delay between the two Raman optical fields scanned over all range between $0$ and $2\pi$, relative to the $\hat{R}_{y}$ reference pulse.
We observe a Ramsey-type fringe with almost perfect visibility of $0.95(2)$ in Fig.~\ref{fig:sfig_ramsey} which along with the Rabi oscillation (Fig.~\ref{fig:fig_gate-char}) provides a clear demonstration of the desired unitary rotations on the atomic spin
$\hat{R}_{\phi}(\theta)=\exp(-i\theta(\cos\phi\,\hat{\sigma_x} + \sin\phi\,\hat{\sigma_y}))$.

\subsubsection*{Transformation of the scattering halo}

This section provides details on the data analysis used in preprocessing the raw data from detector coordinates (position and time-of-flight) to the velocity/momentum coordinates relevant for the physical system.
Since the atoms are in free-fall for the majority of the time from the magnetic trap switch-off (see previous section on experimental procedure), the positions of atoms at the detector essentially correspond to velocities (interchangeable with momentum) \cite{Hodgman2017}.
As seen from Fig.~\ref{fig:sfig_num_mix}(a), the spatial distributions of scattered atoms at the detector is aspherical for $m_J=+1$ (bottom) and spherical for $m_J=0$ (middle).
The deformation in the spatial distribution of the $m_J=+1$ scattering halo from the ideal spherical shell arises due to inhomogeneous forces from the stray magnetic field present in the vacuum chamber during free fall.
Since the accurate determination of the atomic momenta is crucial to identifying the scattered atomic pairs, a distortion correcting shape transform is applied to the raw spatial distribution of atoms ($\mathbf{r}$) to retrieve the momentum distribution ($\mathbf{k}$).

First a spatial distribution (detector coordinate) of atoms in the BECs and the scattering halo are distinguished by the internal state and the new coordinate origin defined at approximately the centre of the corresponding halo.
Background atoms from the BECs, thermal fraction and miscellaneous sources other than the scattered pairs are then removed by only keeping counts lying inside the truncated spherical shell defined by $0.6<r/r_{\textrm{tof}}<1.2$ and $|r_z/r_{\textrm{tof}}|<0.8$, where $r_{\textrm{tof}}\approx25~\textrm{mm}$ is the radius of the scattering halo at the detector.
The resulting $\mathbf{r}$-distribution is then fitted with an ellipsoid which defines the desired smooth shape transform consisting of three orthogonal linear scalings about the centre to reduce each principal axes to unity, which can be suitably identified as the normalised momentum coordinate in the centre of mass reference frame $\mathbf{k}$ (see Fig.~\ref{fig:fig_src-g2}(a)).
A final filter to remove the remaining background restricts the $\mathbf{k}$-distribution to $0.9<k<1.1$ and $|k_z|<0.75$, which corresponds to the truncated momentum space $V$ investigated in this work.
An additional relocation of the coordinate origin by $\vec{\mathcal{K}}$ was however necessary to centre the Gaussian profiles of $g^{(2)}_{ij}(\Delta\mathbf{k})$ at $\Delta\mathbf{k}\approx0$, which is crucial in the implementation of Bell test as it effectively defines the two detection ports in each arm: $(\mathbf{k},\uparrow)$, $(\mathbf{k},\downarrow)$, $(-\mathbf{k},\uparrow)$, $(-\mathbf{k},\downarrow)$ corresponding to the conventional $A+$, $A-$, $B+$, $B-$ detection events, respectively.

\subsubsection*{Correlation functions}

\begin{figure}[htb]
\centering
\includegraphics[width=8cm]{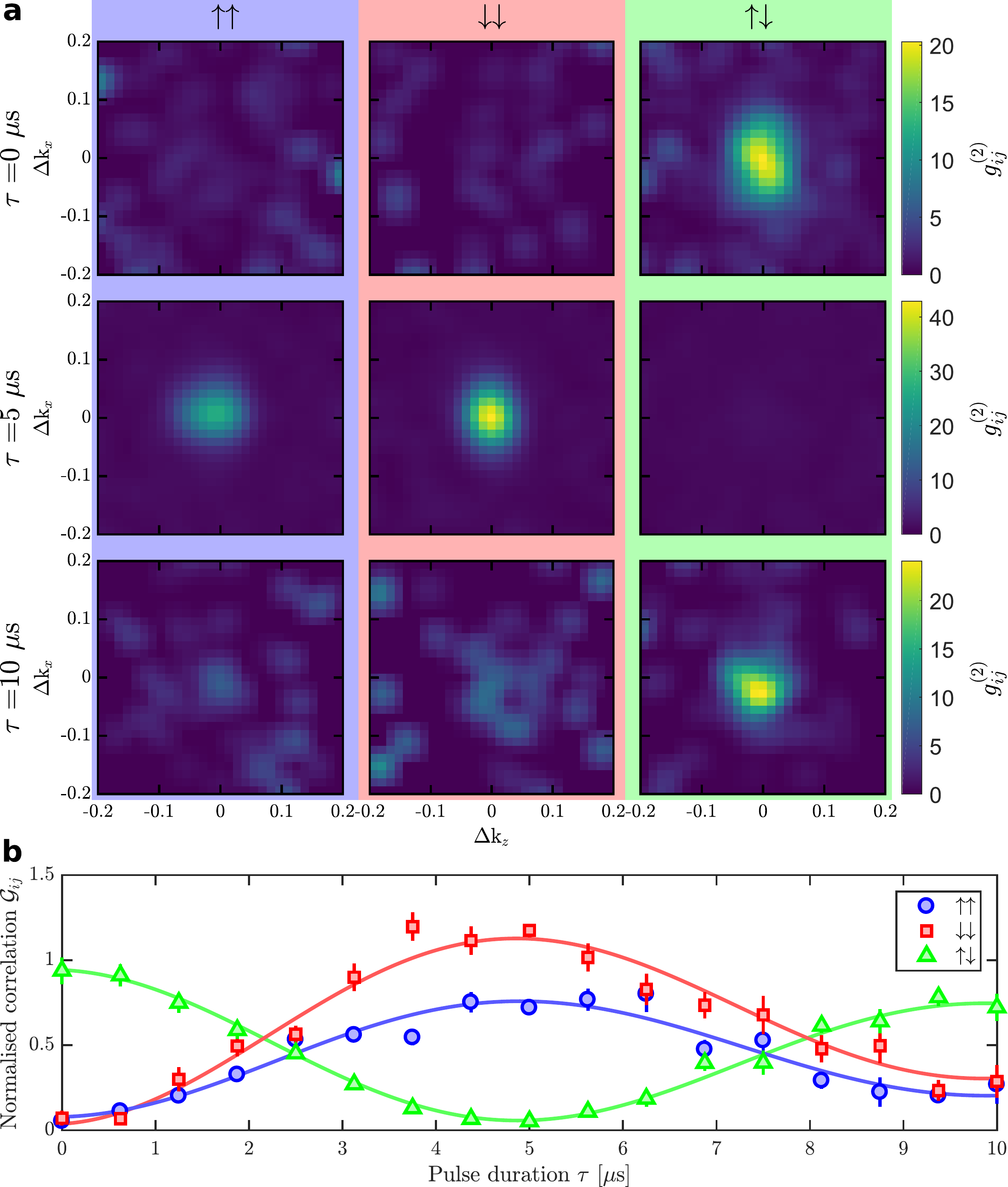}
\caption{\label{fig:sfig_g2_mix}
Evaluated $g^{(2)}$ correlation functions of the scattering halos after various rotations, for pairs with near opposite momenta and various spin-pairing configurations.
(a) $\Delta k_y=0$ slices of $g^{(2)}(\Delta \mathbf{k})$ for various spin rotation pulses (separated by rows) and spin-pairing configurations (separated by columns).
(b) Normalised correlations for the back-to-back condition ($\Delta \mathbf{k}=0$).
The asymmetry in correlation between $\uparrow\uparrow$ (blue circles) and $\downarrow\downarrow$ (red squares) arises from the shape distortion of $\ket{\uparrow}$ atoms' trajectories by stray magnetic fields during time-of-flight.
Error bars correspond to the standard error estimated from bootstrapping.
Solid lines are damped sine curve fits to data.
}
\end{figure}

This section provides details on the two-particle correlation functions in the scattering halos which were ultimately used to construct the $\mathcal{B}$ correlator.
Figure~\ref{fig:sfig_g2_mix}(a) shows the effect of the rotation sequence on the correlation properties of the initially spin and momentum anti-correlated pairs by second-order correlation functions $g^{(2)}$ as defined in Eq.~(\ref{eqn:g2}).
As expected for the $\ket{\Psi^+}$ scattered pairs ($\tau = 0~\mu\textrm{s}$) no correlation is seen between oppositely scattered pairs with the same spin, however with a common $\pi/2$ rotation ($\tau = 5~\mu\textrm{s}$) the spins are always parallel, and finally return to anti-parallel by the nearly spin-flipping $\pi$ pulse ($\tau = 10~\mu\textrm{s}$).
The asymmetry in the Gaussian correlation profiles between different spin types are due to the effects from stray magnetic field during the atoms' free-fall before detection (see previous section for details) which distorts the atomic velocities and thus the correlation profile.
The difference in signal-to-noise ratios of the correlation functions for different rotation sequences are due to the variations in amount of acquired data.

The relative strengths in spin correlations with various rotation pulse durations are summarised in Fig.~\ref{fig:sfig_g2_mix}(b).
Here, the normalised second-order correlation is defined as 
\[\mathcal{G}_{ij} = \dfrac{2g^{(2)}_{ij}}{{\sum_{\alpha,\beta \in \{\uparrow,\downarrow\}} g^{(2)}_{\alpha \beta}}},\]
where the second-order correlation without any argument denotes the value at the exact back-to-back (BB) condition in momentum ($g^{(2)}_{ij} = g^{(2)}_{ij}(\Delta\mathbf{k}\approx0)$).
The BB correlation strengths were determined from a cubic bin of width $0.0138$ in the normalised momentum unit.
As demonstrated in the theory section, the general bipartite correlator can be determined directly from the normalised correlations by Eq.~(\ref{eqn:E_g2}).
The observed asymmetry in the correlation strengths between $\uparrow\uparrow$ and $\downarrow\downarrow$ combinations can be explained by the asymmetry in the correlation volumes due to the shape distortion of the $m_J = 1$ scattering halo, such that the correlation volume integrated strengths recover the expected symmetry.

\subsubsection*{Error analysis}

All statistical uncertainties in the correlation functions, and similarly for other variables, were determined from bootstrapping described here.
From the complete dataset from which the representative correlation function $g$ is determined (see Eq.~(\ref{eqn:g2})), subsets (fractional size $\eta$) are sampled with replacement and analysed identically to produce a distribution of outcomes $G$.
A robust estimate of the standard error of the mean (correlation function in this case) is then given by $\sigma_{\bar{g}} = \sqrt{\eta \cdot \mathrm{Var}(G)}$, independent of the sampling size.

\subsection*{Amplitude criterion for entanglement}
The correlator defined in Eq.~(\ref{eq.corr.1}), assuming local rotations around the $y$-axis by a common angle, can be written as follows
\begin{align}
  \mathcal B(\theta) =\frac12[C_{xx}+C_{zz}+(C_{xx}-C_{zz})\cos2\theta \nonumber\\
  	 +(C_{xz}+C_{zx})\sin2\theta],
\end{align}
where
\begin{align}
	C_{ij}=\av{\hat\sigma_{i}^{(A)}\hat\sigma_{j}^{(B)}}
\end{align}
and $i,j=x,z$. The minimum and the maximum of $\mathcal B(\theta)$ with respect to $\theta$ yield $\mathcal A$  --- the amplitude of oscillations, namely
\begin{align}
  \frac12\left[C_{xx}+C_{zz}-\mathcal A\right]\leqslant\mathcal B(\theta)\leqslant\frac12\left[C_{xx}+C_{zz}+\mathcal A\right],
\end{align}
where
\begin{align}\label{eq.amp}
  \mathcal A=\sqrt{\left(C_{xx}-C_{zz}\right)^2+\left(C_{xz}+C_{zx}\right)^2}.
\end{align}
If the two qubits form a separable (i.e., non-entangled) state, their composite density matrix reads
\begin{align}\label{eq.sep}
  \hat\varrho=\int d\lambda\, p(\lambda)\hat\varrho^{(A)}_\lambda\otimes\hat\varrho^{(B)}_\lambda,
\end{align}
where $p(\lambda)$ is some probability distribution of a variable $\lambda$ and $\hat\varrho^{(i)}(\lambda)$ are the single-qubit density matrices ($i=A,B$).
These single-particle matrices are given by
\begin{align}
  \hat\varrho^{(A)}_\lambda=\frac12(\hat{\mathds1}^{(A)}+\vec a(\lambda)\hat{\vec\sigma}^{(A)}),\ \\ \nonumber \hat\varrho^{(B)}_\lambda=\frac12(\hat{\mathds1}^{(B)}+\vec b(\lambda)\hat{\vec\sigma}^{(B)}).
\end{align}
Here $\hat{\vec\sigma}^{(A/B)}$ is a vector of Pauli matrices for the qubit $A/B$ and $\vec a(\lambda)$ and
$\vec b(\lambda)$ are the corresponding Bloch vectors ($|\vec a(\lambda)|=1$ for pure and 
$|\vec a(\lambda)|<1$ for mixed states for $A$ and analogously for $B$). 

Using the separable state (\ref{eq.sep}) this takes the form
\begin{align}
  C_{ij}=\tr{\hat\varrho\hat\sigma_i^{(A)}\hat\sigma^{(B)}_j}=\int d\lambda\, p(\lambda)a_i(\lambda)b_j(\lambda).
\end{align}
Note that the components of the Bloch vector lying in the $zx$-plane can be parameterised as follows
\begin{align}
  \vec a(\lambda)=\alpha(\lambda)\left(\begin{array}{c}\cos\phi_a^{(\lambda)}\\\sin\phi_a^{(\lambda)}\end{array}\right),\ \ \ 
  \vec b(\lambda)=\beta(\lambda)\left(\begin{array}{c}\cos\phi_b^{(\lambda)}\\\sin\phi_b^{(\lambda)}\end{array}\right),
\end{align}
where $\alpha(\lambda)\leqslant1$ and $\beta(\lambda)\leqslant1$.
With this parameterisation, the amplitude $\mathcal A$ is equal to
\begin{align}
  \mathcal A^2&=(C_{zz}-C_{xx})^2+(C_{zx}+C_{xz})^2\nonumber\\
  	&=\av{\alpha\beta(\sin\phi_a\sin\phi_b-\cos\phi_a\cos\phi_b)}^2\nonumber\\
  	&\quad +\av{\alpha\beta(\cos\phi_a\sin\phi_b+\sin\phi_a\cos\phi_b)}^2\nonumber\\
  &=\av{\alpha\beta\cos(\phi_a+\phi_b)}^2+\av{\alpha\beta\sin(\phi_a+\phi_b)}^2.
\end{align}
Using the Cauchy-Schwarz inequality
\begin{subequations}
  \begin{align}
    &\av{\alpha\beta\cos(\phi_a+\phi_b)}^2\leqslant\av{(\alpha\beta)^2}\av{\cos^2(\phi_a+\phi_b)}\\
    &\av{\alpha\beta\sin(\phi_a+\phi_b)}^2\leqslant\av{(\alpha\beta)^2}\av{\sin^2(\phi_a+\phi_b)}
  \end{align}
\end{subequations}
we obtain
\begin{align}
 & \mathcal A^2 \leqslant \av{(\alpha\beta)^2}\left(\av{\cos^2(\phi_a+\phi_b)}+\av{\sin^2(\phi_a+\phi_b)}\right)\nonumber\\
  	&=\av{(\alpha\beta)^2}\left(\av{\cos^2(\phi_a+\phi_b)+\sin^2(\phi_a+\phi_b)}\right)\nonumber\\
  	&=    \av{(\alpha\beta)^2}\leqslant1.
\end{align}
Therefore $\mathcal A>1$ implies entanglement between the qubits, which was used in Eq.~\ref{eqn:general_Sparam}.

\subsection*{Non-locality}

In our setup, the spin-rotation beams are much larger than the halo size at this point, thus all atoms in the halo are rotated by the same angle $\theta$ and we have access only to the diagonal part of $E$, as $\mathcal B(\theta)=E(\theta,\theta)$.  An extension to implement independent rotations in each atom of the pairs would be experimentally possible, but is beyond the scope of this current work.

Still, using Eq.~(\ref{eq.corr.1}) we can test a wide range of LHV theories. These also take binary outcomes in $A$ and $B$ (i.e., $\uparrow/\downarrow$), but assume that, on average, the results in one of the subsystems 
(either $A$ or $B$) behave like components of a vector, while making no assumptions about the other part. To restrict our system to such an LHV theory, we analyse the properties of rotations in $A$ and $B$ in a dedicated series of experiments, with the results shown in Figs \ref{fig:fig_gate-char} and \ref{fig:sfig_ramsey} demonstrating that they do indeed rotate as vectors. 

Consider two subsystems, where quantities $A$ and $B$ are measured.
The joint probability for observing $A$ and $B$ fulfils the postulates of local realism if
\begin{align}\label{eq.steer}
  P(A,B)=\sum_\lambda p(\lambda)P(A|\lambda)P(B|\lambda),
\end{align}
where $P(A|\lambda)$ or $P(A|\lambda)$ are the conditional probabilities for observing $A$ or $B$ given some value of a hidden variable $\lambda$, governed by the probability distribution $\lambda$.
The conditional probability for observing $B$ given some result $A$ is
\begin{align}\label{eq.pr}
  P(B|A)&=\frac{P(A,B)}{P(A)}\nonumber\\
  &=\sum_\lambda \frac{p(\lambda)P(A|\lambda)}{P(A)}P(B|\lambda)\nonumber\\
  &=\sum_\lambda P(\lambda|A)P(B|\lambda).
\end{align}
Now, imagine that given the quantity measured in $A$, labelled as $J_i^A$, can have binary outcomes for two local settings $i=x,z$, i.e., $J_x^{A}=\pm1$ and $J_z^{A}=\pm1$. 
The two quantities $J_i^{B}$ measured in $B$ are, on average, assumed to be components of a vector of length $1$, which in particular implies
\begin{align}
  -1\leqslant\av{J^B_i}\leqslant 1.
\end{align}
Note that we do not specify how this average is calculated, and the outcomes in $B$ can be binary as well.

The average outcome in $B$, say in the $x$ direction, given the result in $A$ is
\begin{align}
  \av{J^B_x}_A&=\sum_B P(B|A)J^B_x\nonumber\\
  &=\sum_\lambda P(\lambda|A)\sum_BP(B|\lambda)J^B_x,
\end{align}
where
\begin{align}
  -J^B_\lambda\leqslant\sum_BP_Q(B|\lambda)J^B_x\leqslant J^B_\lambda,
\end{align}
and $J^B_\lambda$ is the length of the vector in $B$ given the value of $\lambda$. Thus the upper bound reads
\begin{align}
  \av{J^B_x}_A=\sum_B P(B|A)J^B_x\leqslant\sum_\lambda P(\lambda|A)J^B_\lambda=J^B_A.
\end{align}
We take two orthogonal directions in the $zx$-plane, $\frac1{\sqrt2}(J^B_1-J^B_2)$, and according to the above argument we obtain
\begin{align}\label{eq.ineq}
  -J^B_A\leqslant\frac{\av{J^B_1}_A-\av{J^B_2}_A}{\sqrt 2}\leqslant J^B_A.
\end{align}
Now, we modify the inequality (\ref{eq.ineq}) by multiplying the two averages by the corresponding results in $A$. Using that outcomes in $A$ are binary, we obtain
\begin{align}\label{eq.ineq2}
  -J^B_A\leqslant\frac{J^A_1\av{J^B_1}_A-J^A_2\av{J^B_2}_A}{\sqrt 2}\leqslant J^B_A.
\end{align}
Finally, we average this inequality by the outcomes in $A$. The correlators are equal to
\begin{align}
  \sum_AP(A)J^A_i\av{J^B_i}_A &=\sum_AP(A)J_i^A\sum_B P(B|A)J_i^B\nonumber\\
  &=\sum_{A,B}P(A,B)J_i^AJ_i^B=\av{J_i^AJ_i^B},
\end{align}
while $\sum_AP(A)J^B_A=\av{J^B}\leqslant1$. Thus
\begin{align}
  \left|\av{J^A_1J^B_2}-\av{J^A_2J^B_2}\right|\leqslant\sqrt 2. 
\end{align}
When the system is composed of two qubits, $J^i$ is replaced with a corresponding Pauli operator and the inequality becomes
\begin{align}
  \mathcal S\left(\theta,\theta + \frac{\pi}{2}\right)= |\av{\hat\sigma_1^{(A)}\hat\sigma_1^{(B)}}-\av{\hat\sigma_\perp^{(A)}\hat\sigma_\perp^{(B)}}|\leqslant\sqrt2
\end{align}
for all systems compatible with the LHV model outlined above. To test it, one can analyse the combination of the correlator $\mathcal B$ in Eq.~\ref{eqn:Bell_nonlocal} and plotted in Fig.~\ref{fig:fig_bell}.

\subsubsection*{CHSH inequality for pair-scattering systems}

The most general LHV theory that can be tested with two qubits assumes binary outcomes of the measurements in $A$ and $B$. Both the original Bell inequality~\cite{Bell1964} or its modification proposed by  Clauser, Horne, Shimony and Holt (CHSH)~\cite{clauser1969proposed} use the above assumption.
The CHSH inequality
\begin{equation}\label{eq.chsh}
    \mathrm{B_{\textrm{CHSH}}} = \Big|E(\theta,\phi) + E(\theta',\phi') + E(\theta',\phi) - E(\theta,\phi')\Big|\leqslant 2,
\end{equation}
requires independent rotations in $A$ and $B$ with $\hat{R}^{(A)}_y(\theta) = \exp(-i\theta\hat\sigma_y^{(A)})$ and $\hat{R}^{(B)}_y(\phi) = \exp(-i\phi\hat\sigma_y^{(B)})$ and the measurement of the corresponding correlator
\begin{equation}\label{eq.corr.2}
	E(\theta,\phi) = \av{\hat\sigma_z^{(A)}\hat\sigma_z^{(B)}}_{\theta,\phi}.
\end{equation}

To derive the CHSH inequality for the system where atoms scatter in a pair, we use
the bosonic field operators $\hat\Psi_\alpha(\x)$ for each spin component $\alpha=\pm1,0$.
The Hamiltonian (the summation convention is used) for a $J=1$ BEC reads
\begin{align}
  \hat H&= \int\!\! d^3r\,\bigg[
    \frac{\hbar^2}{2m} \nabla\hat\Psi_\alpha^\dagger(\x) \cdot \nabla\hat\Psi_\alpha(\x) + V(\x) \hat\Psi_\alpha^\dagger(\x)\hat\Psi_\alpha(\x) \nonumber \\
    &+\frac{c_0}{2} \hat\Psi_\alpha^\dagger(\x) \hat\Psi_\beta^\dagger(\x)  \hat\Psi_\beta(\x) \hat\Psi_\alpha(\x) \nonumber \\
    &+\frac{c_1}{2} \hat\Psi_\alpha^\dagger(\x) \hat\Psi_\beta^\dagger(\x)  \mathbf{F}_{\alpha\alpha'} \cdot \mathbf{F}_{\beta\beta'}  
    \hat\Psi_{\beta'}(\x) \hat\Psi_{\alpha'}(\x)\Bigg].
\end{align}
The coefficients $c_{0/1}$ are related to the scattering lengths $a_{0/2}$ in total angular momentum interaction channels $J=0,2$ by $c_0=\frac{4\pi \hbar^2 }{m}
\frac{a_0 + 2 a_2}{3}$ and $c_1=\frac{4\pi \hbar^2 }{m} \frac{a_2 - a_0}{3}$, while $\mathbf{F} = (F_x,F_y,F_z)$ is a vector of spin-1 matrices.

Though the scattering of $m_J=0,1$ pairs from BECs is governed by terms proportional to $c_0$ and~$c_1$, in
the low-density limit spin changing collisions are less probable, and the term proportional to $c_1$ can be safely neglected. Also in this regime, the dynamics of $\hat\Psi_\alpha$ can be found using the Bogoliubov approximation.  Each component is decomposed into the dominant $c$-number term and a quantum
correction, $\hat \Psi_\alpha(\x) =\phi_\alpha(\x)+\hat\delta_\alpha(\x)$. The coherent fields $\phi_\alpha$ are governed by time-dependent Gross-Pitaevskii-type
equations, whereas the fields $\hat \delta_\alpha$ are subject to dynamical Bogoliubov equations which contain both $\hat\delta_\alpha$ and $\hat\delta_\beta^\dagger$ terms. 

Since, in the process of the halo formation, the mean-field energy due to the BECs is small compared to the kinetic energy of the scattered atoms, the only relevant term
proportional to $c_0$ is the production term, and the equation of motion is $i\hbar\partial_t \hat\delta_\alpha=-\frac{\hbar^2\nabla^2}{2m}\hat\delta_{\alpha}+c_0
\phi_{\beta}\phi_\alpha\hat\delta_{\beta}^\dagger$. The presence of the field $\hat\delta_\beta^\dagger$ describes the scattering of atoms from BECs and the
formation of the halo.  The terms proportional to $\phi_{0}^2\hat\delta_{0}^\dagger$ and $\phi_{1}^2\hat\delta_{1}^\dagger$ govern the quantum depletion of the
BECs, irrelevant for the scattering process. As a result, the Bogoliubov equations are symmetric for $\hat \delta_{0/1}$. Therefore, we can consider the same Bell
sequence as in Ref.~\cite{Wasak2018}.

The Bell test starts with the mixing of the two spin components $m_J=0,1$ (from now on denoted as $\downarrow/\uparrow$) independently in two opposite regions of the halo, $A$ and $B$, by the angles
$\phi$ and $\theta$ over the $y$-axis. The many-body angular momentum operators (and the atom-number operators) are
\begin{subequations}
  \begin{eqnarray}
    \hat S_x^{\alpha} &=& \frac12 \int_\alpha\!\!\frac{d\k}{2\pi}\bigg( \hat\delta_{\uparrow}^\dagger(\bk)\hat\delta_{\downarrow}(\bk) + \hat\delta_{\downarrow}^\dagger(\bk)\hat\delta_{\uparrow}(\bk)  \bigg),\label{jx} \\
    \hat S_y^{\alpha} &=& \frac{1}{2i} \int_\alpha\!\!\frac{d\k}{2\pi}\bigg( \hat\delta_{\uparrow}^\dagger(\bk)\hat\delta_{\downarrow}(\bk) - \hat\delta_{\downarrow}^\dagger(\bk)\hat\delta_{\uparrow}(\bk)  \bigg), \label{jy}\\
    \hat S_z^{\alpha} &=& \frac12 \int_\alpha\!\!\frac{d\k}{2\pi}\bigg( \hat\delta_{\uparrow}^\dagger(\bk)\hat\delta_{\uparrow}(\bk) - \hat\delta_{\downarrow}^\dagger(\bk)\hat\delta_{\downarrow}(\bk)  \bigg), \\
    \hat N_{\alpha} &=& \frac12 \int_\alpha\!\!\frac{d\k}{2\pi}\bigg( \hat\delta_{\uparrow}^\dagger(\bk)\hat\delta_{\uparrow}(\bk) + \hat\delta_{\downarrow}^\dagger(\bk)\hat\delta_{\downarrow}(\bk)  \bigg),
  \end{eqnarray}
\end{subequations}
with $\alpha=A,B$.
Finally, a normalised correlator is constructed 
\begin{align}\label{eq.corr}
  E(\theta,\phi)=\frac{\langle\hat S_z^{(A)}\hat S_z^{(B)}\rangle_{\theta,\phi}}{\langle\hat N^{(A)}\hat 	N^{(B)}\rangle_{\theta,\phi}},
\end{align}
where the subscript $\theta,\phi$ denotes averaging over the rotated state.

This correlator $E(\theta,\phi)$ from Eq.~(\ref{eq.corr}) satisfies the Bell inequality of Eq.~(\ref{eq.chsh}) for all LHV theories~\cite{reid1986violations}.
The correlator $E$ can be evaluated analytically in the Bogoliubov theory~\cite{Wasak2018}: $E(\theta,\phi) = -\mathcal{E} \cos(\theta+\phi)$, showing a dependence only in the sum of the angles and oscillations with the amplitude $\mathcal{E} = {(g^{(2)}_{\uparrow\downarrow} - 1)}/{ (g^{(2)}_{\uparrow\downarrow} + 1)}$, which is expressed in terms of the two-particle correlation function:
\begin{equation}
  g^{(2)}_{\uparrow\downarrow}=\frac{\iint\limits_{A\,B} d\k d\k' \av{\hat \delta^\dagger_{\uparrow}(\k)\hat \delta^\dagger_{\downarrow}(\k')\hat \delta_{\downarrow}(\k')\hat \delta_{\uparrow}(\k)}}
  {\iint\limits_{A\,B} d\k d\k'\av{\hat \delta^\dagger_{\uparrow}(\k)\hat \delta^{\phantom\dagger}_{\uparrow}(\k)}\av{\hat \delta^\dagger_{\downarrow}(\k')\hat \delta^{\phantom\dagger}_{\downarrow}(\k')}}.
\end{equation}
Here, the average is calculated in the state prior to rotations. The expression for $\mathrm{B_{\textrm{CHSH}}}$ optimised over angle
settings yields the condition $|\mathcal E| > 1/\sqrt{2}$, which in turn is equivalent to $g^{(2)}_{\uparrow\downarrow}>2\sqrt2 + 3$ for the Bell inequality (\ref{eq.chsh}) to be violated.

The key quantities in the Bell inequality \eqref{eq.chsh} are probed by setting equal angles in the correlator, i.e., $E(\theta,\theta) = \mathcal B(\theta)$.
The observation of $|\mathcal{B}(\theta)| > 1/\sqrt{2}$ signals the detection of the Bell correlations, and the potential for the Bell inequality violation in an actual Bell test experiment with independent settings of the angles in separated regions $A$ and $B$. Finally, we point out that in the low-gain regime, when only a single pair of qubits is scattered, the many-body angular momentum operators are replaced by the Pauli matrices in $A$ and $B$, and the correlator (\ref{eq.corr}) takes the form of Eq.~(\ref{eq.corr.2}). 

\subsubsection*{$\mathcal{B}$ correlator}

The correlation coefficient $E(\theta,\phi)$ given above can be readily evaluated from the single-particle detection resolved in momentum and spin, since
\begin{align}
E(\theta,\phi) &= 
\frac{\sum\nolimits_{\mathbf{k} \in V}\langle\hat S_z^{(\mathbf{k})}\hat S_z^{(-\mathbf{k})}\rangle_{\theta,\phi}}
    {\sum\nolimits_{\mathbf{k} \in V}\langle\hat N^{(\mathbf{k})}\hat N^{(-\mathbf{k})}\rangle_{\theta,\phi}} \nonumber \\
&= \frac	{\expval{\left( \hat{N}_{\uparrow}^{(A)} - \hat{N}_{\downarrow}^{(A)} \right)
 						\left( \hat{N}_{\uparrow}^{(B)} - \hat{N}_{\downarrow}^{(B)} \right)}_{\theta,\phi}}
		{\expval{\left( \hat{N}_{\uparrow}^{(A)} + \hat{N}_{\downarrow}^{(A)} \right)
						\left( \hat{N}_{\uparrow}^{(B)} + \hat{N}_{\downarrow}^{(B)} \right)}_{\theta,\phi}}.
\end{align}
Therefore, it can be expanded in terms of products of number operators across the regions A/B 
\begin{widetext}
\begin{equation}
E = \frac 	{
		\expval{\hat{N}_{\uparrow}^{(A)} \hat{N}_{\uparrow}^{(B)} }
		+ \expval{\hat{N}_{\downarrow}^{(A)} \hat{N}_{\downarrow}^{(B)}}
		- \expval{\hat{N}_{\uparrow}^{(A)} \hat{N}_{\downarrow}^{(B)}}
		- \expval{\hat{N}_{\downarrow}^{(A)} \hat{N}_{\uparrow}^{(B)}}
			}
			{
		\expval{\hat{N}_{\uparrow}^{(A)} \hat{N}_{\uparrow}^{(B)}}
		+ \expval{\hat{N}_{\downarrow}^{(A)} \hat{N}_{\downarrow}^{(B)}}
		+ \expval{\hat{N}_{\uparrow}^{(A)} \hat{N}_{\downarrow}^{(B)}}
		+ \expval{\hat{N}_{\downarrow}^{(A)} \hat{N}_{\uparrow}^{(B)}}
			},
\label{eqn:E_oper_def}
\end{equation}
\end{widetext}
where for convenience the subscript $(\theta,\phi)$ for labelling the general rotated state is assumed for all correlators.
Since the density of the scattering halo is symmetric in momentum ($s$-wave scattering) and spin, $\bar{N} = \expval{\hat{N}_{m}^{(\mathbf{k})}}$ for all $m \in \{\uparrow, \downarrow\}$ and $\mathbf{k} \in V$, observe that each term in Eq.~(\ref{eqn:E_oper_def}) corresponds to a  second-order correlation function
\begin{equation}
\expval{\hat{N}^{(A)}_{i} \hat{N}^{(B)}_{j}} = \bar{N}^2 g^{(2)}_{ij}
\end{equation}
Therefore, the general correlation coefficient $E(\theta,\phi)$ in Eq.~(\ref{eqn:E_oper_def}) can be written in terms of $g^{(2)}$ as
\begin{equation}
E(\theta,\phi) = \frac{g^{(2)}_{\uparrow\uparrow}+g^{(2)}_{\downarrow\downarrow}-g^{(2)}_{\uparrow\downarrow}-g^{(2)}_{\downarrow\uparrow}}
{g^{(2)}_{\uparrow\uparrow}+g^{(2)}_{\downarrow\downarrow}+g^{(2)}_{\uparrow\downarrow}+^{(2)}_{\downarrow\uparrow}}.
\label{eqn:E_g2}
\end{equation}

\end{document}